\documentclass[twocolumn,tighten]{aastex63}
\pdfoutput=1 %for arXiv submission
\usepackage{amsmath,amstext}
\usepackage{apjfonts}
\usepackage{afterpage}
\usepackage{xcolor}
\usepackage{threeparttable}
\usepackage{graphicx}
\usepackage{epsfig}
\usepackage{booktabs}
\usepackage{multirow}
\usepackage{float}
\usepackage{mathrsfs}
\DeclareUnicodeCharacter{1E11}{\k{a}}
\setwatermarkfontsize{100px}

\begin{document}
\author[0009-0003-6348-7143]{Tong Zhao}
\affiliation{School of Physics, Peking University, Beijing 100871, China}

\author[0000-0002-9042-3044]{Renxin Xu}
\affiliation{School of Physics, Peking University, Beijing 100871, China}

\title{A Physically Motivated Compact Parameterization of Neutron Star Atmosphere Emission for Thermal X-ray Pulse-Profile Modeling}

\begin{abstract}
Thermal X-ray pulse-profile modeling depends on the angular and spectral properties of neutron star atmosphere emission. We construct a compact empirical approximation to the specific intensity using eight coefficients for the angular dependence and five for a normalized spectral kernel. The angular prescription is motivated by the formal solution of the radiative-transfer equation, while the spectral normalization preserves a specified bolometric flux. We calibrate this family to numerical atmosphere tables, focusing on return-current-heated and fully ionized hydrogen atmospheres. When incorporated into a semi-analytic pulse-profile model, the approximation reproduces profiles normalized by their phase-averaged flux to within approximately $4\%$ for antipodal hot spots and $5\%$ for non-antipodal hot spots in the tested heated-atmosphere configurations. We also analyze synthetic eXTP observations using different beaming prescriptions to test the conclusion in our previous work based on a linear description of the angular dependence of the emission. The comparisons suggest that the angular dependence of the intensity cannot be simply characterized by a linear function of the cosine value of the emission angle, because its detailed shape and energy dependence will affect the parameter inference. This parameterization can significantly increase the efficiency for analytic and numerical models, and provide a compact framework for examining how atmosphere assumptions affect neutron star pulse profiles and the inferred stellar properties.
\end{abstract}

\section{Introduction}
Measurements of neutron star masses and radii provide a means of constraining the equation of state of matter at densities beyond those accessible in terrestrial experiments. Thermal X-ray pulse-profile modeling offers a particularly valuable approach because the observed modulation carries information about both the stellar spacetime and the emitting surface. As localized hot regions rotate through the observer’s line of sight, gravitational light bending, gravitational redshift, Doppler boosting, and aberration modify the amplitude, shape, and energy dependence of the pulse profile. These effects depend on the stellar mass and radius, but also on the geometry of the emitting regions and the angular distribution of the surface radiation. Extracting reliable constraints on neutron star structure therefore requires a consistent description of both photon propagation and the local emission process~\citep{Beloborodov2002,Pout1,Watts2016}.

Observations with the Neutron Star Interior Composition Explorer (NICER) have demonstrated the potential of this technique for rotation-powered millisecond pulsars. Analyses of PSR J0030+0451 yielded joint constraints on its mass, radius, and surface hot-region geometry, establishing thermal pulse-profile modeling as a practical probe of neutron star structure\citep{Miller2019,Riley2019}. Subsequent analyses of the massive pulsar PSR J0740+6620 extended these measurements to a substantially different mass regime\citep{Miller2021,Riley2021,Choudhury2024,Salmi2024,Salmi2024a,Dittmann2024,Mauviard2025}. These results motivate continued examination of the physical assumptions entering the forward model, particularly those that can change the inferred stellar parameters without being independently constrained by the observations.

An essential ingredient of the forward model is the specific intensity emerging from the neutron star atmosphere. Unlike isotropic blackbody emission, atmospheric radiation generally has a coupled dependence on photon energy and emission angle. The angular emission pattern of the atmosphere is consequently a physical component of the inferred pulse shape, rather than merely a spectrum correction. In commonly adopted atmosphere models, this dependence is specified by the effective temperature, surface gravity and magnetic field, and assumed atmospheric composition. Many NICER analyses have employed nonmagnetic, fully ionized hydrogen atmospheres in which the heat supplying the outgoing radiation is deposited beneath the layers responsible for the emergent spectrum~\citep{Ho2001,Haakonsen2012}. For rotation-powered millisecond pulsars, however, the hot surface regions are thought to be sustained by the bombardment of charged particles returning from the magnetosphere. Whether a deep-heated atmosphere adequately describes this situation depends on where these particles deposit their energy. ~\citet{Baubock2019} showed that the atmospheric structure and radiation pattern depend on the properties of the incident particles and their stopping depths.~\citet{atmosphere} subsequently calculated return-current-heated atmospheres including particle energy deposition and Compton scattering. Their models demonstrated that sufficiently energetic particles can produce emission resembling the deep-heated limit, whereas shallower deposition can generate a hot surface layer and modify both the spectrum and its angular distribution. Depending on the heating conditions and photon energy, the emerging radiation can depart substantially from the usual limb-darkened pattern and can exhibit limb brightening.

The atmospheric angular emission pattern also enters models of thermonuclear (Type~I) X-ray bursts from accreting neutron stars. In burst-oscillation models, a rotating surface brightness asymmetry produces an energy-dependent waveform whose amplitude and harmonic content depend on the local angular emission pattern, making the beaming prescription relevant to mass-radius inference~\citep{Lo2013,Miller2015}. Hot burst atmospheres are often dominated by electron scattering, motivating the use of the corresponding limb-darkening law in waveform calculations. Numerical atmosphere models nevertheless show departures from this law that depend on photon energy and composition, particularly away from the spectral peak and near absorption edges~\citep{Suleimanov2012}. The angular emission pattern also matters when integrating the emission over a rapidly rotating, oblate star: the angular weighting of surface elements affects the observed flux and spectral normalization, and thus enters extensions of the burst cooling-tail method that account for rotation~\citep{Suleimanov2020}. 

However, the specific intensity is calculated using radiative-transfer equations and incorporated into pulse-profile calculations through precomputed tables. Repeated table lookup and interpolation can be computationally expensive. This cost becomes particularly important in GPU-accelerated pulse-profile calculations that reduces a single likelihood evaluation from seconds minutes to millisecond, because the computation of intensity takes up the majority of the computation time~\citep{Zhou2026}. The widespread use of atmosphere models therefore motivates compact representations of the coupled angular and spectral intensity values. 

In this work, we construct a physically motivated parameterized representations of the emergent specific intensity and its angular dependence, which can be calibrated to different atmosphere models, especially for return-current-heated models with different physics parameters. We assess how accurately parameterized beaming functions can reproduce numerical atmosphere calculations and incorporate these functions into a semi-analytic pulse-profile model. We then generate synthetic eXTP-SFA observations and analyze them under different assumptions about the surface emission. By comparing the recovered parameters with their injected values, we examine how an incorrect beaming prescription affects neutron star compactness, viewing geometry, and hot-region properties, and assess the implications for mass-radius inference.

This paper is organized as follows. Section~2 introduces our physically motivated parameterization of the angular and spectral dependence of the emergent intensity. Section~3 presents its calibration to numerical hydrogen atmosphere tables and evaluates the approximation errors. Section~4 compares pulse profiles calculated using the parameterized intensities with those obtained from numerical tables. Section~5 uses synthetic eXTP-SFA observations to investigate how an incorrect beaming prescription affects the inferred stellar and hot-spot parameters. Section~6 summarizes our findings and discusses the limitations of the model. Additional calibration results for helium and partially ionized hydrogen atmospheres are presented in the Appendix.

\section{Physically Motivated Empirical Beaming Functions for X-ray Pulse-Profile Modeling}\label{sec:par}
In our previous work, we used a parameterized function to describe the angular dependence of the emission from the atmosphere~\citep{Zhao2026}, which is a very rough approximation and is only used for synthetic data. Here, we try to find a physically motivated formula to describe the angular distribution of the emergent specific intensity and the spectrum with an analytic family whose coefficients can be calibrated to different atmosphere models. Throughout this section, the $\mu=\cos\alpha$ is evaluated in the local co-moving frame, where $\alpha$ is the angle between the outgoing ray and the surface normal. The effective temperature and local surface gravity are denoted by $T_{\rm eff}$ and $g$, respectively. We fit both the intensity at the surface normal $I(\mu=1)$ and the angular dependence to the numerical result respectively. To separate the angular dependence from the spectral amplitude, we define the angular ratio as
\begin{equation}
 R_I(\mu)
 \equiv \frac{I_\nu(\mu)}{I_\nu(\mu=1)}.
 \label{eq:beaming-rref}
\end{equation}
At fixed energy, the same ratio is obtained whether intensity is expressed per unit frequency or per unit photon energy.

\subsection{The Angular Dependence of Intensity}
For a plane-parallel, optically thick atmosphere with an effective angle-independent source function, the emergent intensity has the formal solution
\begin{equation}
 I_\nu(0,\mu)=\frac{1}{\mu}\int_0^\infty
 S_\nu(\tau_\nu)\exp(-\tau_\nu/\mu)\,d\tau_\nu,
 \qquad \mu>0,
 \label{eq:beaming-formal-solution}
\end{equation}
where $\tau_\nu$ is the extinction optical depth measured inward from the surface. The exponential kernel gives rays at smaller $\mu$ greater sensitivity to shallower layers. Angular beaming therefore encodes the depth dependence of the source function. We approximate this dependence over the layers contributing to the
emergent radiation by
\begin{equation}
 S_\nu(\tau_\nu)\simeq S_{0,\nu}+S_{1,\nu}\tau_\nu^p,
 \qquad S_{0,\nu}\geq0,\quad S_{1,\nu}>0,\quad p>0.
 \label{eq:beaming-source-ansatz}
\end{equation}
Substitution into Eq.~\eqref{eq:beaming-formal-solution} yields
\begin{equation}
 I_\nu(0,\mu)\simeq S_{0,\nu}
       +S_{1,\nu}\Gamma(p+1)\mu^p,
 \label{eq:beaming-source-integral}
\end{equation}
and hence
\begin{equation}
  R_I(\mu)=\frac{K+\mu^p}{1+K},
 \label{eq:beaming-law}
\end{equation}
with
\begin{equation}
 K=\frac{S_{0,\nu}}
 {S_{1,\nu}\Gamma(p+1)\mu^p}.
 \label{eq:beaming-K-interpretation}
\end{equation}
The exponent $p$ describes the effective depth variation of the source contribution, while $K$ measures the depth-independent contribution relative to it. These quantities need not have the same values at different energies or atmospheric parameters. Equation~\eqref{eq:beaming-law} includes the linear law $R_I(\mu)=a+(1-a)\mu$ when $p=1$, with $a=K/(1+K)$, but also allows both concave and convex angular profiles. This flexibility is
motivated by the limited spectral range over which linear beaming is adequate:~\citet{atmosphere} find approximately linear angular dependence near the spectral peak, with departures at lower and higher energies. The source-function argument is an effective description rather than
an exact reduction of the atmosphere calculation. In particular,
scattering can make the source function angle dependent, and Compton
redistribution couples different photon energies.

We define three dimensionless quantities
\begin{equation}
 x=\frac{E}{k_BT_{\rm eff}},\qquad
 \Theta=\frac{k_BT_{\rm eff}}{m_ec^2},\qquad
 \ell=\frac{\kappa_{\rm T}\sigma_{\rm SB}T_{\rm eff}^4}{gc},
 \label{eq:beaming-dimensionless}
\end{equation}
where $\kappa_{\rm T}=0.4\,\mathrm{cm^2\,g^{-1}}$ is the fixed Thomson opacity normalization for the hydrogen models. The variable $x$ measures photon energy relative to the effective thermal scale, whereas $\Theta$ and $\ell$ provide dimensionless coordinates for temperature and gravity.

A useful combination follows from the scaling of non-relativistic free-free absorption at fixed composition,
\begin{equation}
 \kappa_\nu^{\rm ff}\propto
 \rho T^{-1/2}\nu^{-3}
 \left(1-e^{-h\nu/(k_BT)}\right)g_{\rm ff},
 \label{eq:beaming-freefree}
\end{equation}
where $T$ and $\rho$ are the local temperature and density and $g_{\rm ff}$ is the Gaunt factor. Introducing $t=T/T_{\rm eff}$ and $\widehat P=\kappa_{\rm T}P_{\rm gas}/g$, the ideal-gas relation gives
\begin{equation}
 \frac{\kappa_\nu^{\rm ff}}{\kappa_{\rm T}}
 \propto
 \frac{\widehat P\,t^{-3/2}}{\ell\sqrt{\Theta}}
 x^{-3}\left(1-e^{-x/t}\right)g_{\rm ff}.
 \label{eq:beaming-opacity-scaling}
\end{equation}
Constants independent of the atmospheric parameters are suppressed in this proportionality. This scaling motivates the coordinates
\begin{equation}
 z=\frac{\ell}{\ell_*}
       \left(\frac{\Theta}{\Theta_*}\right)^{1/2},
 \qquad
 G=\frac{g}{g_*}
   =\left(\frac{\Theta}{\Theta_*}\right)^4\frac{\ell_*}{\ell}.
 \label{eq:beaming-zG}
\end{equation}
Thus, $z$ is proportional to the inverse of a global free-free opacity pre-factor; it is not itself the full scattering-to-absorption ratio. The local profiles and the Gaunt factor remain in
Eq.~\eqref{eq:beaming-opacity-scaling}. We adopt the fixed reference scales:
\begin{equation}
 \Theta_*=2\times10^{-4},\qquad
 g_*=2\times10^{14}\,\mathrm{cm\,s^{-2}},\qquad
 \ell_*=7.4839\times10^{-6},
 \label{eq:beaming-scales}
\end{equation}
where $\ell_*$ is calculated from $\Theta_*$ and $g_*$ using Eq.~\eqref{eq:beaming-dimensionless}. Equivalently,
\begin{equation}
 z=\left(\frac{k_BT_{\rm eff}}{0.10219979\,\mathrm{keV}}\right)^{9/2}
   \left(\frac{g}{2\times10^{14}\,\mathrm{cm\,s^{-2}}}\right)^{-1}.
 \label{eq:beaming-z-physical}
\end{equation}
With these dimensionless quantities, we find a form of function that can fit the numerical results of different atmosphere models well with only a few free coefficients:
\begin{align}
 R_I(\mu)&=\frac{K+\mu^p}{1+K},\\
 K(x,z,G)&=k_1x^{k_2}z^{k_3}G^{k_4},
 \label{eq:beaming-K-fit}\\
 p(x,z)&=p_1x^{p_2+p_3\ln x}z^{p_4}.
 \label{eq:beaming-p-fit}
\end{align}
This introduces eight coefficients to be fitted to numerical results from the atmosphere models. Taking $k_1$, $p_1>0$ ensures positive $K$ and $p$ without restrictions on the other exponents. The explicit forms in Eqs.~\eqref{eq:beaming-K-fit}--\eqref{eq:beaming-p-fit} are empirical approximations motivated by the transfer and opacity scalings above. Their exponents do not follow uniquely from first principles. The factor $G^{k_4}$ describes residual gravity dependence, since $z$ already depends on $g$.

\subsection{A Compact Approximation to the Emergent Spectrum}
\label{sec:spectral-approximation}
To complement the normalized angular prescription, we construct a common analytic representation of the normal specific intensity. Our aim is to retain the main spectral dependence on photon energy, effective temperature, and surface gravity with few adjustable coefficients. We achieve this by describing a normalized flux spectrum and using the previously calibrated beaming law to determine its conversion to normal intensity. All energies and intensities below refer to the local co-moving frame, and $I_\nu$ and $F_\nu$ are defined per unit frequency.

We use the dimensionless coordinates $x$ and $z$ same as what we used in the angular fit. Collapsing the spectral dependence on temperature and gravity into $z$ is an empirical simplification. It is not an exact reduction of the atmosphere equations. A color-corrected blackbody changes the characteristic energy and amplitude while retaining a fixed Planck shape. To allow departures from this restricted shape, we adopt the generalized kernel
\begin{equation}
 \mathcal{S}(x,z)=\frac{x^n}{\exp[a(z)x^{\beta(z)}]-1},
 \qquad a>0,\quad\beta>0.
 \label{eq:spectrum-kernel}
\end{equation}
The coefficient $a$ sets the energy scale in the exponential, $\beta$ allows its curvature to vary, and $n$ modifies the spectral power index. The kernel contains the Planck shape when $n=3$ and $\beta=1$. More generally, its exponential argument can be written as $x/f_{\rm eff}$, with $f_{\rm eff}=x^{1-\beta}/a$. This provides an effective energy-dependent color factor, allowing the spectrum to depart from a single color temperature. These quantities characterize the emergent spectral shape. They are not identified with a measured temperature or temperature gradient within the atmosphere.

Instead of fitting the amplitude of $\mathcal{S}$ independently, we normalize it analytically. With $s=(n+1)/\beta$, substitution of $t=ax^\beta$ gives
\begin{equation}
 \int_0^\infty\mathcal{S}(x,z)\,dx
 =\frac{\Gamma(s)\zeta(s)}{\beta a^s},\qquad s>1,
 \label{eq:spectrum-integral}
\end{equation}
where $\Gamma$ and $\zeta$ are the Gamma and Riemann zeta functions. The normalized shape is consequently
\begin{equation}
 \Phi(x,z)=\frac{\beta a^s}{\Gamma(s)\zeta(s)}
           \frac{x^n}{\exp(ax^\beta)-1},\qquad
 \int_0^\infty\Phi(x,z)\,dx=1.
 \label{eq:spectrum-phi}
\end{equation}

We adopt the reference flux $F_0=\sigma_{\rm SB}T_{\rm eff}^4$ as the
normalization of the spectral approximation and write
\begin{equation}
 F_\nu=F_0\frac{h}{k_BT_{\rm eff}}\Phi(x,z).
 \label{eq:spectrum-flux}
\end{equation}
The factor $h/(k_BT_{\rm eff})=dx/d\nu$ ensures that $\int F_\nu\,d\nu=F_0$. This normalization convention was held fixed during calibration. Its identification with the physical bolometric flux requires $T_{\rm eff}$ to be the total-flux effective
temperature.\footnote{In an externally heated atmosphere the emergent bolometric flux includes both intrinsic and deposited heating contributions. If a table instead labels its temperature by the heating contribution alone, $F_0$ is the heating flux scale, and the total-flux interpretation requires the intrinsic
contribution to be included.}

To minimize the number of adjustable coefficients, we represent the remaining atmospheric dependence by two power laws,
\begin{equation}
 a(z)=a_0z^{a_z},\qquad \beta(z)=b_0z^{b_z}.
 \label{eq:spectrum-shape-powers}
\end{equation}
The five spectral coefficients are thus $a_0$, $a_z$, $b_0$, $b_z$, and $n$, with $a_0,b_0>0$. These power laws are empirical choices selected for their compactness and their performance on both grids. For $n=3$, $\beta=1$, and $a=1/f_c$, the normalized spectrum reduces to a diluted blackbody with color temperature $f_cT_{\rm eff}$ and dilution factor $f_c^{-4}$. The additional shape freedom is retained to describe departures from that limit.

The outward spectral flux then satisfies
\begin{equation}
 \begin{split}
 F_\nu&=2\pi\int_0^1 I_\nu(\mu)\mu\,d\mu\\
 &=\pi I_\nu(1)\frac{K+2/(p+2)}{1+K}.
 \end{split}
 \label{eq:spectrum-angular-integral}
\end{equation}
It follows that
\begin{equation}
 I_\nu(1)=\frac{F_\nu}{\pi}Q,\qquad
 Q=\frac{1+K}{K+2/(p+2)}.
 \label{eq:spectrum-Q}
\end{equation}
The factor $Q$ is determined entirely by the existing angular coefficients. It retains the energy and atmospheric dependence of $K(x,z,G)$ and $p(x,z)$.

Combining Eqs.~\eqref{eq:spectrum-flux} and \eqref{eq:spectrum-Q} gives the normal-intensity approximation:
\begin{equation}
\frac{I_\nu(1)}{T_{\rm eff}^3}
 =C\frac{\pi^4}{15}Q\Phi(x,z),\qquad
 C=\frac{2k_B^3}{h^2c^2}.
\label{eq:spectrum-normal-intensity}
\end{equation}
The corresponding intensity at arbitrary emission angle is
\begin{equation}
 I_\nu(\mu)=C T_{\rm eff}^3\frac{\pi^4}{15}\Phi(x,z)
 \frac{K+\mu^p}{K+2/(p+2)}.
 \label{eq:spectrum-full-intensity}
\end{equation}
This construction introduces five new spectral coefficients in addition to the eight angular coefficients, for a total of thirteen per atmosphere model. The angular integral and spectral normalization are exact identities for the adopted analytic functions. The choice of spectral kernel and the fitted dependence of its shape parameters remain phenomenological, rather than solutions of the full radiative-transfer problem.

\section{Calibration to Numerical Atmosphere Tables}
We calibrated this same functional family to five different atmosphere models given in~\citet{Salmi2023}. In this section, we show the results of the return-current-heated hydrogen atmosphere with the particle energy distribution power-law index $\delta=2$, and the fully-ionized hydrogen atmosphere denoted as NSX~\citep{Salmi2023,Ho2001}. The heated model belongs to the class considered by~\citet{atmosphere}. Other fit results are shown in the Appendix. We choose a proper domain of parameters suitable for neutron star thermal X-ray pulse profiles. The selected domain is listed in Table~\ref{table:1}.
\begin{table}[t]
\caption{Domain of Parameters}
\begin{center}
\begin{tabular}{c c}
\toprule
Parameters & Range\\
\midrule
Effective Temperature &$0.02~\mathrm{keV}\leq k_BT_{\rm eff}\leq0.45~\mathrm{keV}$\\ 
Photon Energy & $0.1~\mathrm{keV}\leq E\leq3~\mathrm{keV}$ \\
x & $\leq30$\\
Gravitational Acceleration & $13.7~\mathrm{cm\,s^{-2}}\leq\log_{10}g\leq14.6~\mathrm{cm\,s^{-2}}$ \\
\bottomrule
\end{tabular}
%\end{threeparttable}
\end{center}
\label{table:1}
\end{table}

These restrictions were applied jointly. The common gravity interval was used even though the NSX source table extends to higher gravities. The retained heated and NSX tables contain 9,030 and 17,530 angular curves, respectively, with 11 and 67 directions per curve. The fit results for the heated atmosphere and the NSX table are shown in Table~\ref{table:2}.

\begin{table}[tb]
 \centering
 \caption{Coefficients of the common beaming family for the return-current-heated atmosphere and the NSX atmosphere.}
 \label{table:2}
 \begin{tabular}{lrr}
  \hline
  Parameter & Heated atmosphere & NSX \\
  \hline
  $k_1$ & $6.12192783\times10^{-4}$ & $4.80592285\times10^{-4}$ \\
  $k_2$ & $1.91037621$ & $2.05620112$ \\
  $k_3$ & $0.390481506$ & $0.395265253$ \\
  $k_4$ & $0.0269532436$ & $0.0478843034$ \\
  $p_1$ & $0.189587940$ & $0.305522834$ \\
  $p_2$ & $0.798876627$ & $0.443586021$ \\
  $p_3$ & $-0.0354961294$ & $0.0312071340$ \\
  $p_4$ & $-0.00664998522$ & $-0.000790744942$ \\
  \hline
 \end{tabular}
\end{table}

\begin{figure}
\centering
\includegraphics[width=8.5cm]{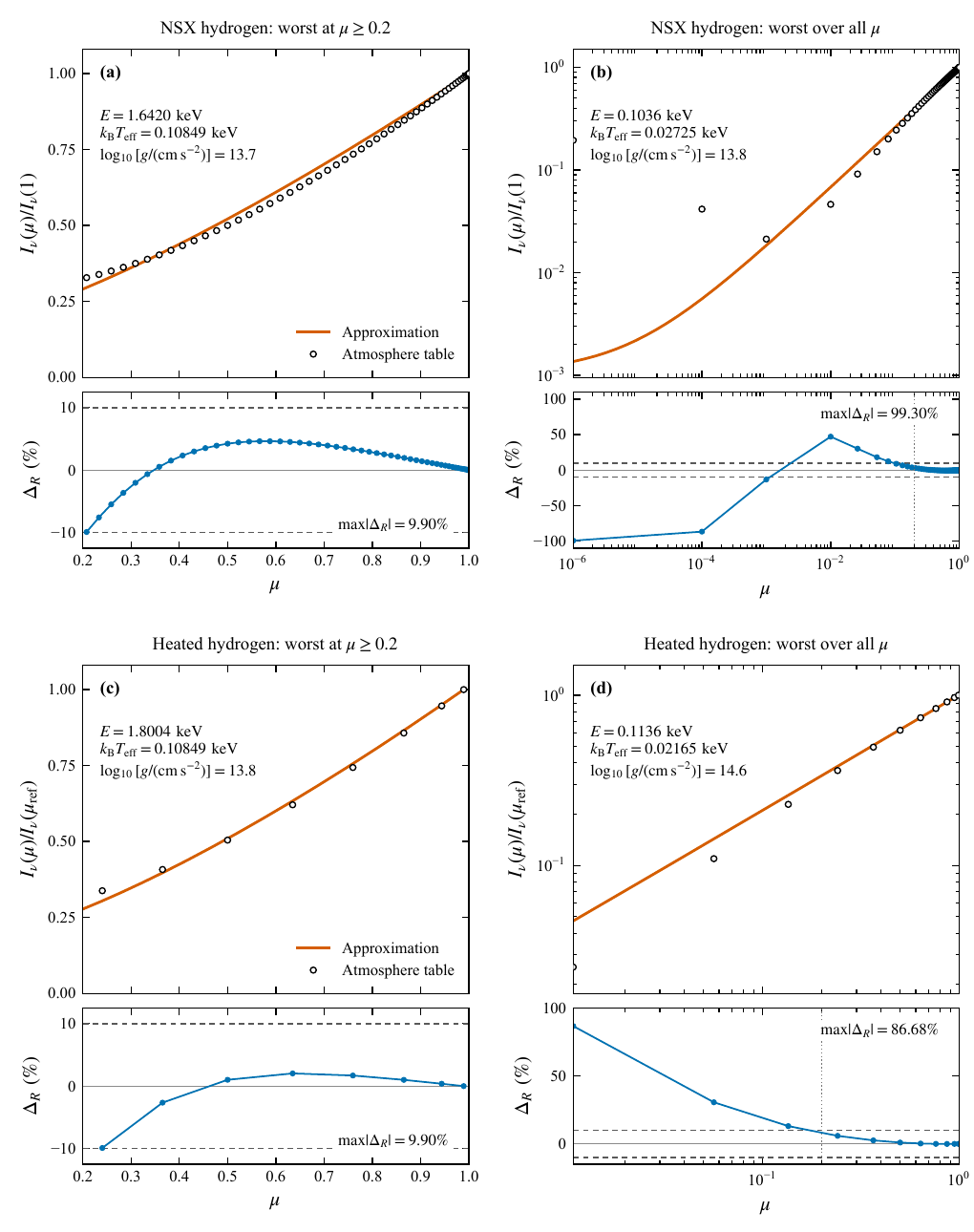}
\caption{Comparison of the analytic beaming approximation with the NSX hydrogen atmosphere (top row) and heated hydrogen atmosphere (bottom row). The left panels show the curves containing the largest absolute relative errors among tabulated samples with $\mu\geq0.2$; the right panels show the curves containing the largest errors over the full sampled angular range. The photon energy, effective temperature, and surface gravity are indicated in each panel. Open circles represent the atmosphere tables, and solid orange curves represent the analytic approximation. The angular ratio is normalized at $\mu=1$. The lower portion of each panel shows the signed relative residual, $\Delta_R=100(R_{\rm fit}/R_{\rm tab}-1)$, with dashed horizontal lines marking $\pm10\%$. The largest errors for $\mu\geq0.2$ are approximately $-9.90\%$ for both models. Over all sampled angles, the largest errors are $-99.30\%$ at $\mu=10^{-6}$ for NSX and $+86.68\%$ at $\mu=0.010886$ for the heated atmosphere. Vertical dotted lines in the right-hand residual panels indicate $\mu=0.2$, below which the fitting constraint was relaxed. Note the different axis scales between panels.}
\label{fig:Angular}
\end{figure}

Defining the relative reconstruction error by
\begin{equation}
 \epsilon_\nu(\mu)=
 \left|\frac{R_{\nu}^{\rm fit}(\mu)}
 {R_{\nu}^{\rm table}(\mu)}-1\right|,
 \label{eq:beaming-error}
\end{equation}
the point-wise maximum errors over all selected tabulated points with $\mu\geq0.2$ are $9.90\%$ for the heated atmosphere table and $9.90\%$ for the NSX atmosphere table. The corresponding point-wise median errors are $0.917\%$ and $0.506\%$. These medians depend on the different angular sampling of the two tables. Comparison of the analytic beaming approximation with the numerical results of the NSX hydrogen atmosphere and the heated hydrogen atmosphere are shown in Fig~\ref{fig:Angular}. The left panels show the curves containing the largest absolute relative errors among tabulated samples with $\mu\geq0.2$; the right panels show the curves containing the largest errors over the full sampled angular range. We can see that the larger discrepancies occur at grazing angles. Because for $K,p>0$, Eq.~\eqref{eq:beaming-law} is monotonic in $\mu$ and cannot reproduce an angular upturn. This design is suitable for the thermal X-ray pulse-profile modeling because when the light from a hot spot is at the grazing angle, the observed flux is dominated by the contribution from the other hot spot, or the total flux is weak and leads to a small signal to noise ratio. The linear fit made by~\citet{atmosphere} followed the same trend, but even our worst fit result is better than the linear fit.

For comparisons intended to isolate the angular distribution, the quantity held fixed must also be specified. If the local angle-integrated flux $F_\nu$ is held fixed, a flux-preserving construction is
\begin{equation}
 I_\nu(\mu)=\frac{F_\nu}{\pi}
 \frac{ K+\mu^p}{ K+2/(p+2)},
 \label{eq:beaming-fixed-flux}
\end{equation}
which satisfies $2\pi\int_0^1I_\nu(\mu)\mu\,d\mu=F_\nu$. Equation~\eqref{eq:beaming-fixed-flux} is an algebraic normalization of
the adopted angular law, not an additional atmosphere-spectrum fit. The resulting local intensity can be supplied to the pulse-profile calculation, with the spectral prescription treated separately.

We fitted the same spectral form to the heated hydrogen atmosphere with $\delta=2$ and fixed incident-particle and intrinsic-flux parameters, and to the supplied hydrogen-atmosphere grid denoted NSX. The angular coefficients were kept at their previously calibrated values. The heated and NSX grids provide 9,030 and 17,530 normalization intensities, respectively. Table~\ref{tab:spectrum-coefficients} lists the fitted coefficients, and each set applies globally across its atmosphere grid.

\begin{table}[tb]
 \centering
 \caption{Five spectral coefficients of the common normalized generalized Planck form. The eight previously fitted angular coefficients are held fixed. Values are rounded for presentation.}
 \label{tab:spectrum-coefficients}
 \begin{tabular}{lrr}
  \hline
  Parameter & Heated atmosphere & NSX \\
  \hline
  $a_0$ & $1.528042591$ & $1.888249476$ \\
  $a_z$ & $-0.134679290$ & $-0.103688833$ \\
  $b_0$ & $0.647876944$ & $0.597660199$ \\
  $b_z$ & $0.0668454494$ & $0.0528951391$ \\
  $n$ & $2.705726540$ & $2.905924750$ \\
  \hline
 \end{tabular}
\end{table}

We quantify normal-intensity reconstruction by
\begin{equation}
 \epsilon_\nu=
 \left|\frac{I_\nu^{\rm fit}(1)}{I_\nu^{\rm table}(1)}-1\right|,
 \label{eq:spectrum-error}
\end{equation}
Table~\ref{tab:spectrum-errors} summarizes the errors over all retained normalization points. Relative errors of $I_\nu$ and $I_\nu/T_{\rm eff}^3$ are identical at fixed temperature.

\begin{table}[tb]
 \centering
 \caption{Normal-intensity reconstruction errors on the calibration
 grids. Percentiles are taken over individual energy--temperature--
 gravity points.}
 \label{tab:spectrum-errors}
 \begin{tabular}{lrr}
  \hline
  Statistic & Heated atmosphere & NSX \\
  \hline
  Median & $4.220\%$ & $3.666\%$ \\
  90th percentile & $9.181\%$ & $7.434\%$ \\
  99th percentile & $12.653\%$ & $10.404\%$ \\
  Maximum & $14.813\%$ & $11.873\%$ \\
  Points above $10\%$ & $647/9030$ & $282/17530$ \\
  \hline
 \end{tabular}
\end{table}

\begin{figure}
\centering
\includegraphics[width=8.5cm]{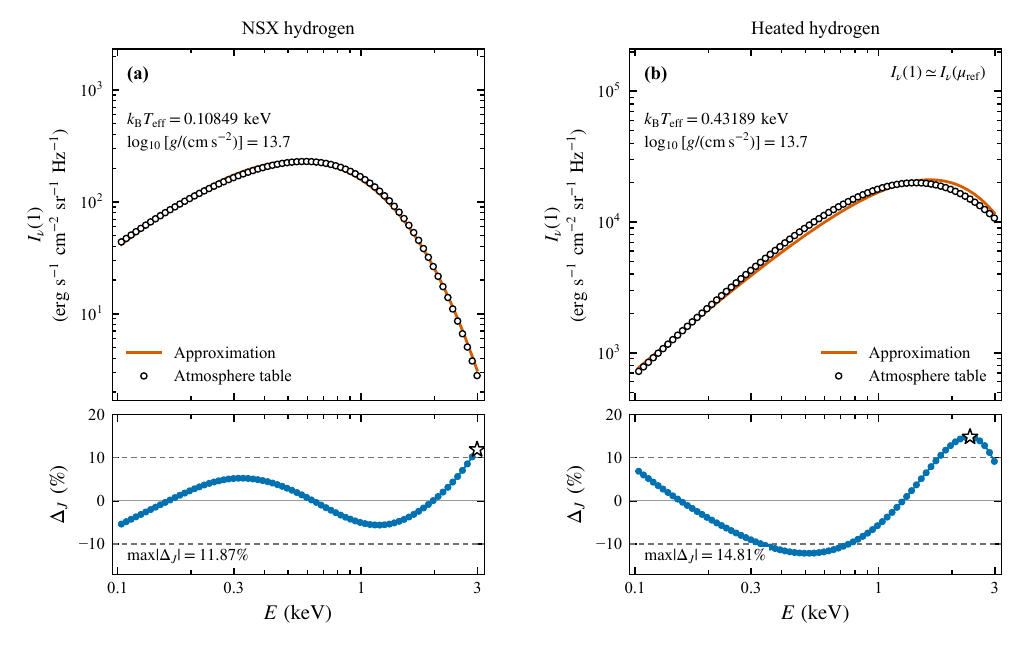}
\caption{Comparison of the analytic normal-intensity spectra with the NSX hydrogen atmosphere (a) and heated hydrogen atmosphere (b). For each model, the spectrum shown contains the largest absolute relative error across the retained calibration grid. Open circles denote the tabulated specific intensity per unit frequency, $I_\nu(1)$, and solid orange curves show the analytic approximation. The effective temperatures are indicated in the panels; both spectra have $g=5.012\times10^{13}\,\mathrm{cm\,s^{-2}}$. For the heated atmosphere, the tabulated intensity at $\mu_{\rm ref}=0.989114$ is used as a proxy for $I_\nu(1)$. The lower panels show the signed relative residuals, $\Delta_J=100[I_{\nu,\rm fit}/I_{\nu,\rm tab}-1]$, with horizontal dashed lines marking $\pm10\%$. Stars identify the largest absolute residuals: $+11.87\%$ at $E=2.988$~keV for NSX and $+14.81\%$ at $E=2.373$~keV for the heated atmosphere. These relative residuals are unchanged when the intensities are divided by $T_{\rm eff}^{3}$.}
\label{fig:Spectrum}
\end{figure}

We also show the worst fit results of the spectrum in Fig~\ref{fig:Spectrum}. Larger discrepancies usually occur far from the spectrum peak, or at very large or small temperature.

\begin{figure}
\centering
\includegraphics[width=8.5cm]{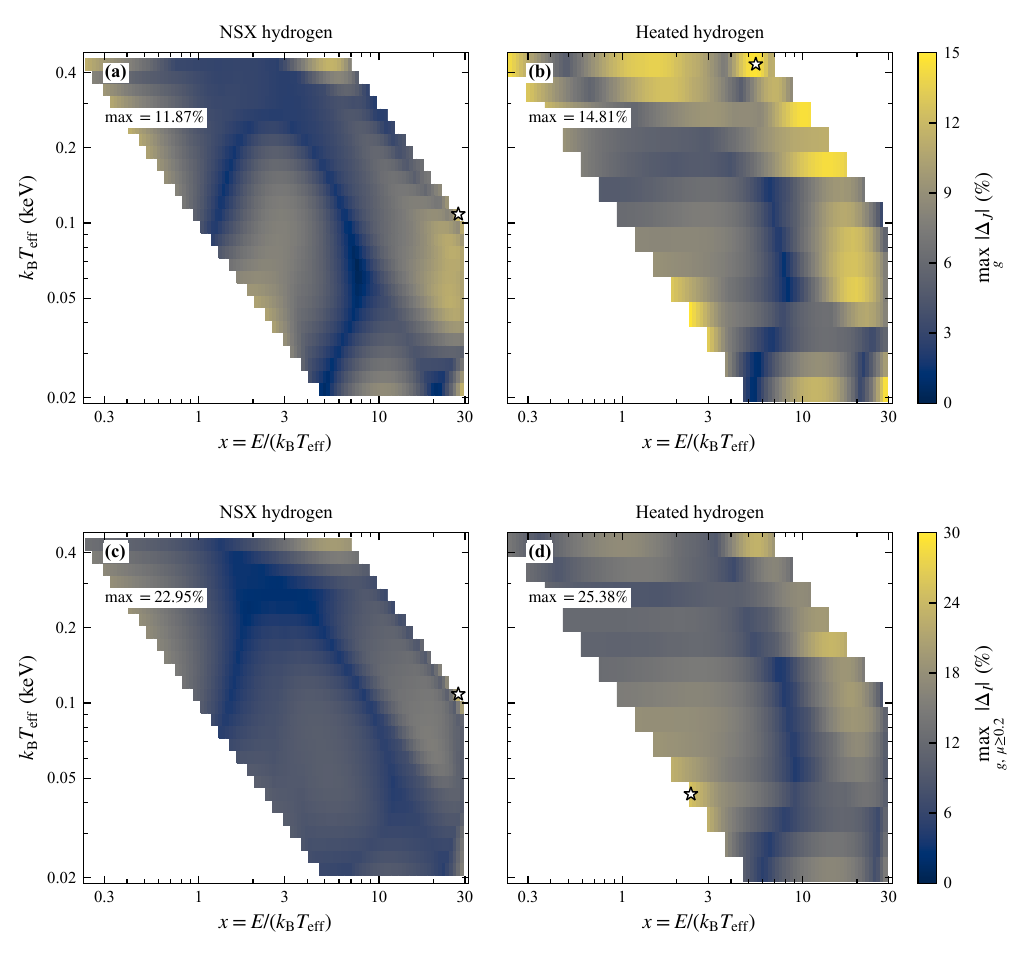}
\caption{Relative-error maps of the analytic intensity approximation for the NSX hydrogen atmosphere (left column) and heated hydrogen atmosphere (right column), evaluated across the retained grids in $x=E/(k_{\rm B}T_{\rm eff})$ and $T_{\rm eff}$. The top row shows $\max_g|\Delta_J|$, where $J=I_\nu(1)/T_{\rm eff}^{3}$ and $\Delta_J=100(J_{\rm fit}/J_{\rm tab}-1)$. For the heated atmosphere, the tabulated intensity at $\mu_{\rm ref}=0.989114$ is used as a proxy for $I_\nu(1)$. The bottom row shows $\max_{g,\,\mu\geq0.2}|\Delta_I|$, where $\Delta_I=100[I_{\nu,\rm fit}(\mu)/I_{\nu,\rm tab}(\mu)-1]$, for the combined spectral and angular approximation. The combined intensity uses an angular factor normalized at exactly $\mu=1$. Each colored cell represents a retained grid location, with maxima taken over the available gravities and, in the bottom row, the tabulated angles satisfying $\mu\geq0.2$. White regions contain no retained data, and stars identify the global maxima. The maximum normal-intensity errors are $11.87\%$ and $14.81\%$ for NSX and the heated atmosphere, respectively; the corresponding combined-intensity maxima are $22.95\%$ and $25.38\%$. The panels within each row share a color scale, while the scales differ between rows. These are reconstruction errors on the calibration grids.}
\label{fig:Map}
\end{figure}

The relative-error maps of the analytic intensity approximation evaluated across the retained grids is shown in Fig.~\ref{fig:Map}. The bottom row shows the relative-error of the combined spectral and angular approximation. The maximum point-wise relative-error can reach $25\%$ (labeled by stars), but it happens only at the edge of the retained grids region and at small $\mu$. If we choose a narrow range of the data to do the fit, the maximum relative-error will decrease further.

The advantage of our analytic presentation over a simple interpolation formula is that some coefficients also have physical meaning. For example, for the NSX atmosphere, $n=2.9$ while $n=2.7$ for the return-current-heated atmosphere. This means the spectrum of the NSX atmosphere is closer to a blackbody spectrum with $n=3$. This is reasonable because NSX atmosphere is equivalent to a deep-heated atmosphere where the local thermal dynamic equilibrium is satisfied. The larger $k_4$ of the NSX shows that it is more sensitive to the gravity.

\section{Comparison of Pulse Profiles Generated by Numerical Intensity Tables and the Fitted Approximate Beaming Functions}

The point-wise relative-error map do not directly specify the error in an observed pulse profile. Each spot samples a phase-dependent range of emission angles and local photon energies, and its contribution is weighted by the projected area and relativistic transfer factors. Contributions from different spots can also dilute or reinforce the intensity residuals. We therefore compare pulse profiles obtained by inserting the numerical intensities and the fitted intensity approximation into the same flux calculation.

We use the small-spot flux equation of\citet{Zhao2026}, based on the Schwarzschild plus Doppler (S+D) approximation and the analytic light-bending relation. This model assume that each spot is small, circular, emitting photons from its center and has a uniform temperature distribution. $\theta$ denotes the observer inclination. For spot $i$, $\zeta_i$ is the spot colatitude and $T_i$ is the effective temperature. The ratio of the area of spot 2 to the area of spot 1 is defined as $a_r$. The observed total flux is the sum of the two spot contributions,
\begin{equation}
 \mathcal{F}_{E_\infty}(\phi)
 =\mathcal{F}_{E_\infty,1}(\phi_1)
 +\mathcal{F}_{E_\infty,2}(\phi_2).
 \label{eq:pulse-total}
\end{equation}
Choosing the first spot to have zero longitude at $\phi=0$, we set $\phi_1=\phi$ and $\phi_2=\phi+\pi+\Delta\phi$. Thus, $\Delta\phi=0$ denotes a longitudinal separation of $180^\circ$. Thus antipodal spots require $\zeta_2=\pi-\zeta_1$, $\Delta\phi=0$, $a_r=1$ and $T_1=T_2=T_{eff}$. We choose the return-current-heated atmosphere model for the intensity. In the approximate-intensity calculation, we use Equation~\eqref{eq:spectrum-full-intensity} at $(E_i',\mu_i',T_i,g_i)$. Since that expression gives intensity per unit frequency, it is converted to intensity per unit photon energy using $I'_{E'}=I'_{\nu'}/h$, with consistent energy units. In the table calculation, we interpolate the numerical intensity at the same local arguments. The geometry and transfer prescription are common to the two calculations. Both the spectral and angular approximations are therefore tested together. Propagation time delays and stellar oblateness are omitted. Because these choices are shared by both intensity prescriptions, their difference measures the effect of replacing the numerical atmosphere intensities within the approximate parameterized functions. It does not measure the accuracy of the approximation of the pulse-profile model itself.

\begin{figure*}[t]
 \centering
 \includegraphics[width=\textwidth]{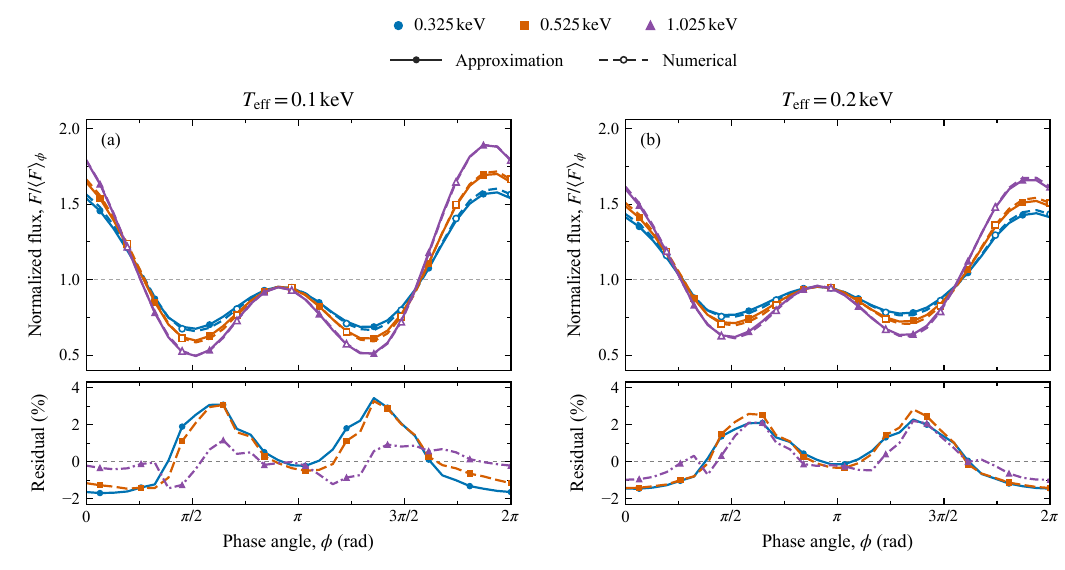}
 \caption{Comparison of normalized pulse profiles calculated with the fitted intensity approximation (solid curves) and the numerical return-current-heated hydrogen intensity table (dashed curves), for two equal-area antipodal spots. Both spots have $k_BT_{\rm eff}=0.1$~keV in panel~(a) and $0.2$~keV in panel~(b). Blue circles, orange squares, and purple triangles denote observed energies of $0.325$, $0.525$, and $1.025$~keV, respectively. Each curve is divided by its own phase-averaged flux. The lower panels show $\Delta_{\rm prof}$ from Equation~\eqref{eq:pulse-residual}, in percent.}
 \label{fig:pulse-antipodal}
\end{figure*}

The common settings inferred from the accompanying research code are the neutron star mass $M=1.5 M_{solar}$, radius $R=12$~km, and the spin frequency $f=300$~Hz. The displayed profiles are normalized separately at each observed energy by their own phase-averaged flux:
\begin{equation}
 f_{E_\infty}(\phi)=
 \frac{\mathcal{F}_{E_\infty}(\phi)}
 {\langle\mathcal{F}_{E_\infty}\rangle_\phi}.
 \label{eq:pulse-normalization}
\end{equation}
Here the phase average is taken over a complete rotation. The signed residual in the lower panels is
\begin{equation}
 \Delta_{\rm prof}(E_\infty,\phi)=100\left[
 \frac{f^{\rm app}_{E_\infty}(\phi)}
 {f^{\rm tab}_{E_\infty}(\phi)}-1\right].
 \label{eq:pulse-residual}
\end{equation}
Figure~\ref{fig:pulse-antipodal} shows two antipodal spots viewed at $\theta=70^\circ$, with $\zeta_1=50^\circ$, $\zeta_2=130^\circ$, and $\Delta\phi=0$. The spots have equal areas and the same effective temperature: $k_BT_1=k_BT_2=0.1$~keV in panel~(a) and $0.2$~keV in panel~(b). We compare the pulse profiles at observed photon energies $E_\infty=0.325$, $0.525$, and $1.025$~keV. The displayed shape residuals remain within approximately $4\%$. The residuals are phase dependent, with their larger positive excursions near the pulse minima at the two lower energies. Thus, the largest local intensity errors at grazing emission angles do not translate directly into errors of the same magnitude in these pulse profiles. Projected-area weighting suppresses emission close to the visibility boundary, and emission from the other spot can further reduce its fractional contribution. This suppression depends on the spot geometry and relative brightness.

\begin{figure*}[t]
 \centering
 \includegraphics[width=\textwidth]{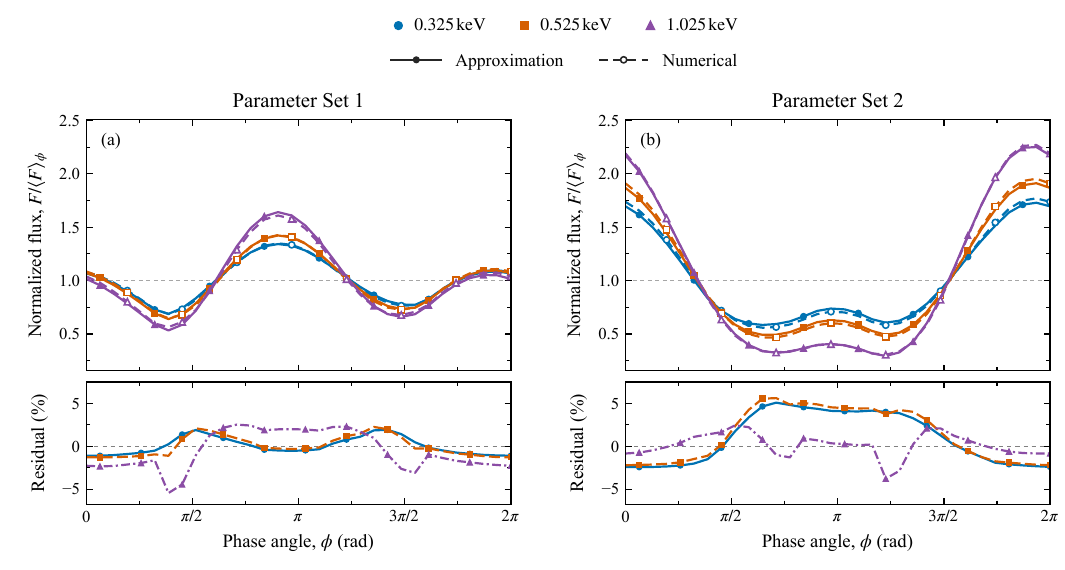}
 \caption{Comparison of normalized pulse profiles for two non-antipodal spots using the numerical return-current-heated hydrogen intensities and the fitted approximation. Panels~(a) and~(b) correspond to parameter sets 1 and 2 in Table~\ref{tab:nonantipodal-parameters}. The secondary spot longitude is $\phi+\pi+\Delta\phi$, and $a_r$ is the secondary-to-primary area ratio. Colors, symbols, line styles, normalization, and residuals are the same as in Figure~\ref{fig:pulse-antipodal}. The lower panels display shape discrepancies of up to approximately $5\%$ for the selected examples.}
 \label{fig:pulse-nonantipodal}
\end{figure*}

To examine more geometry configurations, we allow the two spots to have independent colatitudes, temperatures, and areas. We define $a_r=S_2^\prime/S_1'$ and retain the phase-offset convention $\phi_2=\phi+\pi+\Delta\phi$. The two parameter sets used in Figure~\ref{fig:pulse-nonantipodal} are listed in Table~\ref{tab:nonantipodal-parameters}. Temperatures in the table are local effective temperatures and photon energies in the figure are measured at the observer.

\begin{table}[tb]
 \centering
 \caption{Non-antipodal spot parameters for Figure~\ref{fig:pulse-nonantipodal}.}
 \label{tab:nonantipodal-parameters}
 \begin{tabular}{lrr}
  \hline
  Parameter & Set 1 & Set 2 \\
  \hline
  $\theta$ (deg) & 60 & 80 \\
  $\zeta_1$ (deg) & 40 & 55 \\
  $\zeta_2$ (deg) & 135 & 115 \\
  $k_BT_1$ (keV) & 0.15 & 0.25 \\
  $k_BT_2$ (keV) & 0.20 & 0.15 \\
  $a_r=S_2'/S_1'$ & 1.2 & 1.5 \\
  $\Delta\phi$ (deg) & $+5$ & $-10$ \\
  \hline
 \end{tabular}
\end{table}

The asymmetric configurations produce different peak heights and energy dependences because the relative contribution of each spot changes with both energy and phase. The fitted intensities reproduce these features, with the displayed residuals remaining within approximately $5\%$.

\section{Impact of an Incorrect Angular Emission Model}
In our previous works, we found that if the beaming function can be simply approximated by a linear function of $\mu$ as suggested in~\citet{atmosphere}, fitting a more isotropic emitter with a more strongly normal-peaked beaming law can underestimate compactness for two antipodal spots~\citep{Zhao2025}. This behavior involves correlated changes in compactness and viewing geometry. 

Now, with our more general atmosphere approximation that can accurately reproduce a particular numerical table. In this section, we examine this limitation using the curved angular prescription calibrated to the return-current-heated and the deep-heated NSX atmosphere tables. In general, both atmospheres have energy-dependent spectra and beaming, and their relative anisotropy can vary with temperature, gravity, and emission angle. To separate the angular effect from differences in the atmosphere spectra, we use the same blackbody spectral flux in both the generating and fitted models. For atmosphere label $m\in\{\mathrm{heated},\mathrm{NSX}\}$, the local intensity is
\begin{equation}
 I_{E',m}'(\mu)=B_{E'}(T_{\rm eff})\,
 \frac{K_m+\mu^{p_m}}{K_m+2/(p_m+2)},
 \label{eq:beaming-control}
\end{equation}
with $K_m$ and $p_m$ evaluated at $(E',T_{\rm eff},g)$ using the coefficients in Table~\ref{table:2}. This expression is normalized by the angle-integrated spectral flux,
\begin{equation}
 2\pi\int_0^1 I_{E',m}'(\mu)\mu\,d\mu
 =\pi B_{E'}(T_{\rm eff}).
 \label{eq:wrong-beaming-flux}
\end{equation}
Thus, changing $m$ changes the angular distribution while retaining the same local spectral-flux prescription at a given effective temperature. This construction is a diagnostic emission model. It does not include the different spectra of the two self-consistent atmospheres.

In this section, we generate synthetic pulse profiles for a neutron star with antipodal hot spots. The stellar mass is fixed at 1.5 solar masses and we consider spin frequencies of 10 and 300~Hz to investigate the role of rotational Doppler effects. The proposed injections use two equal-area, equal-temperature antipodal spots and the return-current-heated atmosphere with $\delta=2$. The return-current-heated atmosphere is used to generate the reference profiles. A fit with the same atmosphere provides a recovery check, and a fit with NSX intensities measures the effect of the atmosphere mismatch. Injected and best-fit parameter values are listed in Table~\ref{table:fit}. To avoid working with very small numerical values, we define fiducial values for the spot area $dS$ and distance $D$. They are $dS_0=10^6$~m$^2$ and $D_0=1.2$~kpc. Then, we define a dimensionless factor $A=(dS/D^2)/(dS_0/D_0^2)$ as a free parameter in the model. The best-fit value is the point corresponding to the posterior maximum.

\begin{table*}[t]
\centering
\caption{Injected values and fit results.
Results are given as best-fit values with upper and lower uncertainties.}
\label{table:fit}

\small
\setlength{\tabcolsep}{4pt}
\renewcommand{\arraystretch}{1.2}

\begin{tabular*}{\textwidth}{@{\extracolsep{\fill}}lccccc@{}}
\toprule
 & & \multicolumn{2}{c}{NSX}
   & \multicolumn{2}{c}{Return-current heated} \\
\cmidrule(lr){3-4} \cmidrule(lr){5-6}
Parameter & Injected
 & $10$~Hz & $300$~Hz
 & $10$~Hz & $300$~Hz \\
\midrule
$R$ (km) & 12
 & $11.963^{+1.185}_{-0.663}$
 & $11.317^{+0.070}_{-0.048}$
 & $13.322^{+1.271}_{-1.342}$
 & $11.927^{+0.308}_{-0.154}$ \\

$\theta$ (deg) & 65
 & $69.026^{+4.673}_{-8.994}$
 & $58.520^{+2.888}_{-3.218}$
 & $72.627^{+3.261}_{-17.669}$
 & $59.867^{+6.375}_{-8.949}$ \\

$\zeta$ (deg) & 50
 & $43.866^{+11.514}_{-7.603}$
 & $57.492^{+3.110}_{-3.078}$
 & $38.434^{+22.236}_{-5.875}$
 & $56.023^{+8.301}_{-7.537}$ \\

$T_{\rm eff}$ (keV) & 0.2
 & $0.201^{+0.003}_{-0.004}$
 & $0.204^{+0.000}_{-0.000}$
 & $0.196^{+0.004}_{-0.003}$
 & $0.200^{+0.001}_{-0.001}$ \\

$A$ & 1
 & $1.024^{+0.123}_{-0.072}$
 & $0.951^{+0.009}_{-0.007}$
 & $1.127^{+0.127}_{-0.131}$
 & $0.991^{+0.031}_{-0.016}$ \\
\bottomrule
\end{tabular*}
\end{table*}

For an eXTP-SFA realization, the unnormalized spectral energy flux, including interstellar absorption, is converted to a photon flux and folded through the eXTP instrumental response. The neutral-hydrogen column density is fixed at $N_{\rm H}=1.5\times10^{20}~{\rm cm}^{-2}$, and we assume an exposure time of $10^6$~s. We consider photon energies between 0.3 and 1.5~keV. For computational efficiency, this range is divided into 24 energy bins of width 0.05~keV, while each rotational cycle is divided into 32 phase bins. With the resulting number of counts per energy-phase bin, the Poisson counting uncertainty can be well approximated by a Gaussian uncertainty. The instrumental background obtained from the eXTP simulation is added when constructing the synthetic data and subsequently subtracted in the fitting procedure~\citep{Qi2023}. The likelihood is therefore written as
\begin{equation}
{\cal L}({\cal D}|{\cal M},\Theta)=\exp(-\frac{1}{2}\chi^2)=\exp{[-\frac{1}{2}\sum_i\frac{({\cal M}_i-{\cal D}_i)^2}{\sigma_i^2}]},
\end{equation}
where ${\cal M}_i$ and ${\cal D}_i$ denote the model-predicted and synthetic photon counts, respectively, in the $i$-th energy-phase bin, and $\sigma_i$ is the corresponding uncertainty.

The parameter inference has large uncertainties when the frequency is 10~Hz and the atmosphere model is the return-current-heated atmosphere. This is because the constrain on the radius arises from both rotational Doppler effects and the angular anisotropy of the emergent intensity. When the Doppler effect is negligible and the beaming is isotropic, the model has an intrinsic degeneracy that can lead to large uncertainties in the parameter inference~\citep{Zhao2026}. The results also show that even if we consider only the angular dependence of the real atmosphere model, our conclusion based on linear approximated angular dependence $R_I=(1+h\mu)/(1+h)$ does not capture the parameter biases produced by realistic atmosphere beaming.  Thus, although the linear prescription adopted in~\citet{atmosphere,Zhao2025} may reproduce the angular dependence of the specific intensity with relatively small deviations, these deviations can propagate non-trivially through the pulse-profile model and bias the inferred neutron-star parameters.

\section{Summary}
The proposed family separates a normalized angular distribution from a compact spectral description. The formal transfer solution motivates the angular form, and the spectral kernel admits an analytic normalization. This makes the roles of angular beaming and spectral shape explicit with thirteen coefficients per atmosphere. The fitted exponents and their dependence on $x$, $z$, and $G$ remain empirical. The construction is therefore a calibrated representation of numerical intensities, rather than an independent atmosphere calculation.

There are several physical limits to the present representation. For $K,p>0$, the angular law is monotonic and cannot describe limb brightening or an angular turnover. Grazing-angle errors can be large even when the contribution to a particular pulse profile is small. That contribution must be checked for the actual emitting geometry, especially when only one spot dominates. In addition, the fits cover the retained table points rather than the full range of possible atmosphere parameters. The larger residuals for partially ionized hydrogen in the Appendix show that the same compact family does not provide uniform accuracy across compositions and ionization states. Applications requiring smaller errors may need a restricted calibration domain, a more flexible functional form, or direct numerical intensities.

Our final section demonstrates that the effect of the angular emission pattern on pulse-profile parameter inference is non-trivial. The effect cannot be characterized solely by the overall strength of the angular anisotropy. It also depends on the detailed shape and energy dependence of the angular emission pattern and on its correlations with the stellar and geometrical parameters. A more comprehensive assessment will therefore require exploration of a broader parameter space, including more complex spot geometries and additional atmosphere models.

\appendix
We apply the same eight-coefficient angular prescription and five-coefficient spectral kernel to the supplied helium and partially ionized hydrogen atmosphere tables. Table~\ref{tab:additional-coefficients} gives the resulting coefficients in the notation of Equations~\eqref{eq:beaming-K-fit}, \eqref{eq:beaming-p-fit}, and~\eqref{eq:spectrum-shape-powers}. The dimensionless coordinates retain the reference scales in Equation~\eqref{eq:beaming-scales} and the explicit definition of $z$ in Equation~\eqref{eq:beaming-z-physical}. In particular, the opacity normalization used to define these coordinates is kept fixed when applying the helium coefficients. It should not be replaced by the physical helium Thomson opacity without transforming or refitting the coefficients. This fixed reference is a coordinate convention, not an assertion that hydrogen and helium have the same opacity.

\begin{table}[h!]
 \centering
 \caption{Fitted coefficients for the helium and partially ionized hydrogen tables.}
 \label{tab:additional-coefficients}
 \begin{tabular}{lrr}
  \hline
  Parameter & Helium & Partially ionized H \\
  \hline
  $k_1$ & $2.03148532406\times10^{-4}$ & $6.83472531540\times10^{-4}$ \\
  $k_2$ & $2.01355679500$ & $1.70497650144$ \\
  $k_3$ & $0.457943538395$ & $0.438772035924$ \\
  $k_4$ & $0.0507257417565$ & $0.389133719806$ \\
  $p_1$ & $0.254473447497$ & $0.121060072056$ \\
  $p_2$ & $0.657140848858$ & $1.12004279746$ \\
  $p_3$ & $-0.0192068059378$ & $-0.0719254741514$ \\
  $p_4$ & $0.000516659318662$ & $0.00599873454424$ \\
  \hline
  $a_0$ & $1.58386913805$ & $1.89199788982$ \\
  $a_z$ & $-0.113678215923$ & $-0.145967898177$ \\
  $b_0$ & $0.613571180822$ & $0.613348859502$ \\
  $b_z$ & $0.0516908447097$ & $0.0726672966091$ \\
  $n$ & $2.64164937488$ & $2.92381659715$ \\
  \hline
 \end{tabular}
\end{table}

\begin{figure*}[t]
 \centering
 \includegraphics[width=\textwidth]{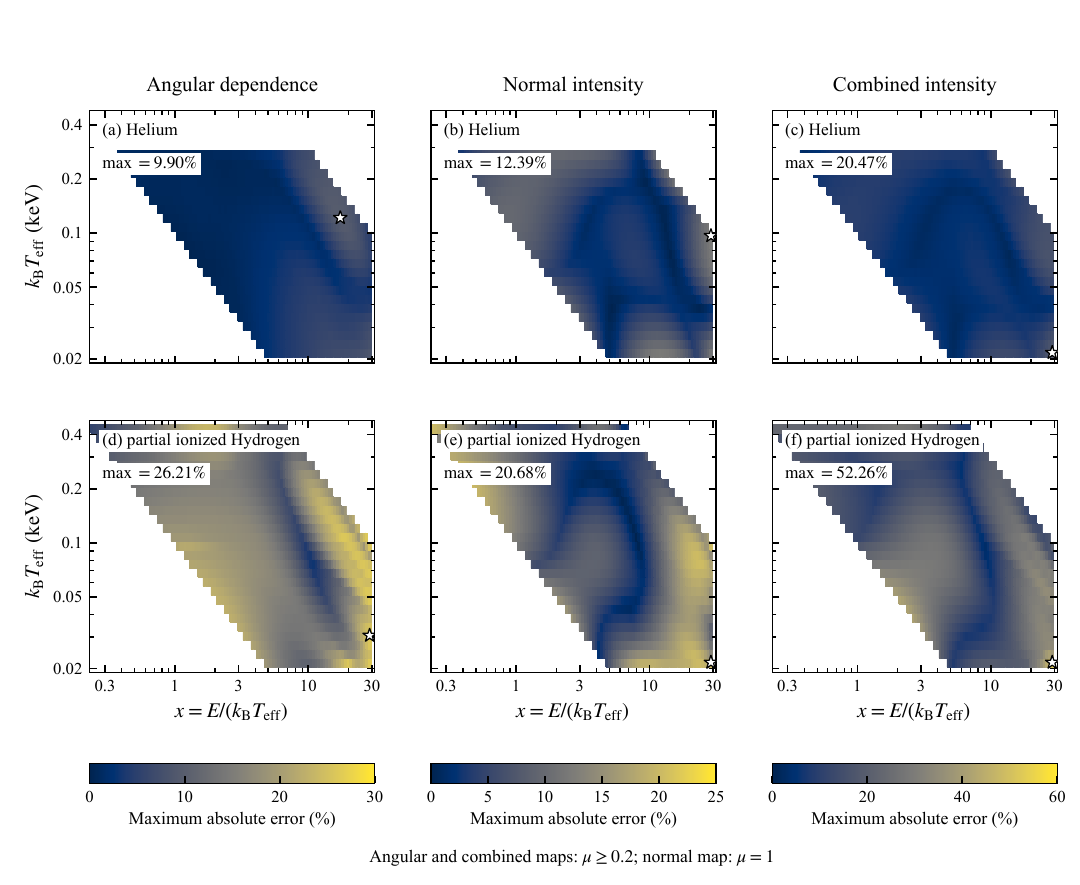}
 \caption{Point-wise relative-error maps for the fitted helium atmosphere (top row) and partially ionized hydrogen atmosphere (bottom row). From left to right, the panels show the maximum absolute relative errors in the angular ratio, the normal intensity, and the combined intensity at each retained $(x,T_{\rm eff})$ location. Maxima are taken over the available gravities, and additionally over angles with $\mu\geq0.2$ for the angular and combined maps. Stars identify the global maxima, whose values are printed in the panels and listed in Table~\ref{tab:additional-errors}. Each column shares a color scale between the two atmospheres. Blank regions contain no retained table points.}
 \label{fig:additional-error-maps}
\end{figure*}

Figure~\ref{fig:additional-error-maps} separates the angular, normal-intensity, and combined-intensity reconstruction errors. Each map reports the largest point-wise error over the available gravities and, where relevant, the retained emission angles. The plotted domain is the intersection of the selection in Table~\ref{table:1} with each source grid.

\begin{table}[h!]
 \centering
 \caption{Maximum reconstruction errors in Figure~\ref{fig:additional-error-maps}. Angular and combined errors are restricted to $\mu\geq0.2$.}
 \label{tab:additional-errors}
 \begin{tabular}{lrr}
  \hline
  Quantity & Helium & Partially ionized H \\
  \hline
  Angular ratio & $9.90\%$ & $26.21\%$ \\
  Normal intensity & $12.39\%$ & $20.68\%$ \\
  Combined intensity & $20.47\%$ & $52.26\%$ \\
  \hline
 \end{tabular}
\end{table}

The helium fit has an accuracy comparable to the hydrogen fits in the main text, with a maximum angular error of $9.90\%$ and a maximum combined-intensity error of $20.47\%$. The partially ionized hydrogen fit is less accurate: its maximum angular, normal-intensity, and combined errors are $26.21\%$, $20.68\%$, and $52.26\%$, respectively. These values are summarized in Table~\ref{tab:additional-errors}. The broad residual structure in the partially ionized maps indicates a limitation of the adopted global functional form. The free-free scaling used to motivate the coordinates does not by itself describe the additional opacity and ionization dependence of this atmosphere. However, choosing a smaller domain within the numerical table can decrease the fitting error. Especially, our global functional form fails when fitted to the carbon atmosphere, because it has absorption edges and the beaming behavior is also completely different. We believe that our global functional can also fit well to different heated atmosphere models given in~\citet{atmosphere}. However, the data is not available.

Spectral and angular residuals can reinforce or partly cancel at a given point. Their separate global maxima need not occur at the same atmospheric parameters or emission angle, so they cannot simply be added to obtain the combined maximum. The errors reported here characterize reconstruction on the retained calibration grids. They do not establish the accuracy of helium or partially ionized hydrogen pulse profiles, which have not been tested in this work.

\bibliographystyle{apj}
\bibliography{MS}

@ARTICLE{Qi2023,
       author = {{Qi}, Liqiang and {Li}, Gang and {Xu}, Yupeng and {Zhang}, Juan and {Ge}, Mingyu and {Xiao}, Jingyu and {Ye}, Wentao and {Xiao}, Yunxiang and {Li}, Xiaobo},
        title = "{Application of the observation simulator in the eXTP mission}",
      journal = {Experimental Astronomy},
         year = 2023,
        month = aug,
       volume = {56},
       number = {2-3},
        pages = {517-536},
          doi = {10.1007/s10686-023-09910-y},
       adsurl = {https://ui.adsabs.harvard.edu/abs/2023ExA....56..517Q}
}

@ARTICLE{Haakonsen2012,
       author = {{Haakonsen}, Christian Bernt and {Turner}, Monica L. and {Tacik}, Nick A. and {Rutledge}, Robert E.},
        title = "{The McGill Planar Hydrogen Atmosphere Code (McPHAC)}",
      journal = {\apj},
         year = 2012,
        month = apr,
       volume = {749},
       number = {1},
          eid = {52},
        pages = {52},
          doi = {10.1088/0004-637X/749/1/52},
       adsurl = {https://ui.adsabs.harvard.edu/abs/2012ApJ...749...52H}
}

@ARTICLE{Ho2001,
       author = {{Ho}, Wynn C.~G. and {Lai}, Dong},
        title = "{Atmospheres and spectra of strongly magnetized neutron stars}",
      journal = {\mnras},
         year = 2001,
        month = nov,
       volume = {327},
       number = {4},
        pages = {1081-1096},
          doi = {10.1046/j.1365-8711.2001.04801.x},
archivePrefix = {arXiv},
       eprint = {astro-ph/0104199},
 primaryClass = {astro-ph},
       adsurl = {https://ui.adsabs.harvard.edu/abs/2001MNRAS.327.1081H}
}

@ARTICLE{Watts2016,
       author = {{Watts}, Anna L. and {Andersson}, Nils and {Chakrabarty}, Deepto and {Feroci}, Marco and {Hebeler}, Kai and {Israel}, Gianluca and {Lamb}, Frederick K. and {Miller}, M. Coleman and {Morsink}, Sharon and {{\"O}zel}, Feryal and {Patruno}, Alessandro and {Poutanen}, Juri and {Psaltis}, Dimitrios and {Schwenk}, Achim and {Steiner}, Andrew W. and {Stella}, Luigi and {Tolos}, Laura and {van der Klis}, Michiel},
        title = "{Colloquium: Measuring the neutron star equation of state using x-ray timing}",
      journal = {Reviews of Modern Physics},
         year = 2016,
        month = apr,
       volume = {88},
       number = {2},
          eid = {021001},
        pages = {021001},
          doi = {10.1103/RevModPhys.88.021001},
archivePrefix = {arXiv},
       eprint = {1602.01081},
 primaryClass = {astro-ph.HE},
       adsurl = {https://ui.adsabs.harvard.edu/abs/2016RvMP...88b1001W}
}

@ARTICLE{Zhao2026,
       author = {{Zhao}, Tong and {Psaltis}, Dimitrios and {{\"O}zel}, Feryal and {Beklen}, Elif},
        title = "{Uncovering Correlations and Biases in Parameter Inference from Neutron Star Pulse-profile Modeling}",
      journal = {\apj},
         year = 2026,
        month = jul,
       volume = {1006},
       number = {1},
          eid = {34},
        pages = {34},
          doi = {10.3847/1538-4357/ae77ed},
archivePrefix = {arXiv},
       eprint = {2412.12283},
 primaryClass = {astro-ph.HE},
       adsurl = {https://ui.adsabs.harvard.edu/abs/2026ApJ..1006...34Z}
}

@ARTICLE{Zhao2025,
       author = {{Zhao}, Tong and {Psaltis}, Dimitrios and {{\"O}zel}, Feryal},
        title = "{A Counterintuitive Correlation between Neutron Star Radii Inferred from Pulse Modeling and Surface Emission Beaming}",
      journal = {\apj},
         year = 2025,
        month = apr,
       volume = {982},
       number = {2},
          eid = {112},
        pages = {112},
          doi = {10.3847/1538-4357/adb9ea},
archivePrefix = {arXiv},
       eprint = {2412.12284},
 primaryClass = {astro-ph.HE},
       adsurl = {https://ui.adsabs.harvard.edu/abs/2025ApJ...982..112Z}
}

@ARTICLE{Dittmann2024,
       author = {{Dittmann}, Alexander J. and {Miller}, M. Coleman and {Lamb}, Frederick K. and {Holt}, Isiah M. and {Chirenti}, Cecilia and {Wolff}, Michael T. and {Bogdanov}, Slavko and {Guillot}, Sebastien and {Ho}, Wynn C.~G. and {Morsink}, Sharon M. and {Arzoumanian}, Zaven and {Gendreau}, Keith C.},
        title = "{A More Precise Measurement of the Radius of PSR J0740+6620 Using Updated NICER Data}",
      journal = {\apj},
         year = 2024,
        month = oct,
       volume = {974},
       number = {2},
          eid = {295},
        pages = {295},
          doi = {10.3847/1538-4357/ad5f1e},
archivePrefix = {arXiv},
       eprint = {2406.14467},
 primaryClass = {astro-ph.HE},
       adsurl = {https://ui.adsabs.harvard.edu/abs/2024ApJ...974..295D}
}

@ARTICLE{Choudhury2024,
       author = {{Choudhury}, Devarshi and {Salmi}, Tuomo and {Vinciguerra}, Serena and {Riley}, Thomas E. and {Kini}, Yves and {Watts}, Anna L. and {Dorsman}, Bas and {Bogdanov}, Slavko and {Guillot}, Sebastien and {Ray}, Paul S. and {Reardon}, Daniel J. and {Remillard}, Ronald A. and {Bilous}, Anna V. and {Huppenkothen}, Daniela and {Lattimer}, James M. and {Rutherford}, Nathan and {Arzoumanian}, Zaven and {Gendreau}, Keith C. and {Morsink}, Sharon M. and {Ho}, Wynn C.~G.},
        title = "{A NICER View of the Nearest and Brightest Millisecond Pulsar: PSR J0437{\textendash}4715}",
      journal = {\apjl},
         year = 2024,
        month = aug,
       volume = {971},
       number = {1},
          eid = {L20},
        pages = {L20},
          doi = {10.3847/2041-8213/ad5a6f},
archivePrefix = {arXiv},
       eprint = {2407.06789},
 primaryClass = {astro-ph.HE},
       adsurl = {https://ui.adsabs.harvard.edu/abs/2024ApJ...971L..20C}
}

@ARTICLE{Salmi2024,
       author = {{Salmi}, Tuomo and {Deneva}, Julia S. and {Ray}, Paul S. and {Watts}, Anna L. and {Choudhury}, Devarshi and {Kini}, Yves and {Vinciguerra}, Serena and {Cromartie}, H. Thankful and {Wolff}, Michael T. and {Arzoumanian}, Zaven and {Bogdanov}, Slavko and {Gendreau}, Keith and {Guillot}, Sebastien and {Ho}, Wynn C.~G. and {Morsink}, Sharon M. and {Cognard}, Isma{\"e}l and {Guillemot}, Lucas and {Theureau}, Gilles and {Kerr}, Matthew},
        title = "{A NICER View of PSR J1231{\ensuremath{-}}1411: A Complex Case}",
      journal = {\apj},
         year = 2024,
        month = nov,
       volume = {976},
       number = {1},
          eid = {58},
        pages = {58},
          doi = {10.3847/1538-4357/ad81d2},
archivePrefix = {arXiv},
       eprint = {2409.14923},
 primaryClass = {astro-ph.HE},
       adsurl = {https://ui.adsabs.harvard.edu/abs/2024ApJ...976...58S}
}

@ARTICLE{Lo2013,
       author = {{Lo}, Ka Ho and {Miller}, M. Coleman and {Bhattacharyya}, Sudip and {Lamb}, Frederick K.},
        title = "{Determining Neutron Star Masses and Radii Using Energy-resolved Waveforms of X-Ray Burst Oscillations}",
      journal = {\apj},
         year = 2013,
        month = oct,
       volume = {776},
       number = {1},
          eid = {19},
        pages = {19},
          doi = {10.1088/0004-637X/776/1/19},
archivePrefix = {arXiv},
       eprint = {1304.2330},
 primaryClass = {astro-ph.HE},
       adsurl = {https://ui.adsabs.harvard.edu/abs/2013ApJ...776...19L}
}

@ARTICLE{Mauviard2025,
       author = {{Mauviard}, Lucien and {Guillot}, Sebastien and {Salmi}, Tuomo and {Choudhury}, Devarshi and {Dorsman}, Bas and {Gonz{\'a}lez-Caniulef}, Denis and {Hoogkamer}, Mariska and {Huppenkothen}, Daniela and {Kazantsev}, Christine and {Kini}, Yves and {Olive}, Jean-Francois and {Stammler}, Pierre and {Watts}, Anna L. and {Mendes}, Melissa and {Rutherford}, Nathan and {Schwenk}, Achim and {Svensson}, Isak and {Bogdanov}, Slavko and {Kerr}, Matthew and {Ray}, Paul S. and {Guillemot}, Lucas and {Cognard}, Isma{\"e}l and {Theureau}, Gilles},
        title = "{A NICER View of the 1.4 M$_{{\ensuremath{\odot}}}$ Edge-on Pulsar PSR J0614-3329}",
      journal = {\apj},
         year = 2025,
        month = dec,
       volume = {995},
       number = {1},
          eid = {60},
        pages = {60},
          doi = {10.3847/1538-4357/ae145d},
archivePrefix = {arXiv},
       eprint = {2506.14883},
 primaryClass = {astro-ph.HE},
       adsurl = {https://ui.adsabs.harvard.edu/abs/2025ApJ...995...60M}
}

@ARTICLE{Salmi2024a,
       author = {{Salmi}, Tuomo and {Choudhury}, Devarshi and {Kini}, Yves and {Riley}, Thomas E. and {Vinciguerra}, Serena and {Watts}, Anna L. and {Wolff}, Michael T. and {Arzoumanian}, Zaven and {Bogdanov}, Slavko and {Chakrabarty}, Deepto and {Gendreau}, Keith and {Guillot}, Sebastien and {Ho}, Wynn C.~G. and {Huppenkothen}, Daniela and {Ludlam}, Renee M. and {Morsink}, Sharon M. and {Ray}, Paul S.},
        title = "{The Radius of the High Mass Pulsar PSR J0740+6620 With 3.6 Years of NICER Data}",
      journal = {arXiv e-prints},
         year = 2024,
        month = jun,
          eid = {arXiv:2406.14466},
        pages = {arXiv:2406.14466},
          doi = {10.48550/arXiv.2406.14466},
archivePrefix = {arXiv},
       eprint = {2406.14466},
 primaryClass = {astro-ph.HE},
       adsurl = {https://ui.adsabs.harvard.edu/abs/2024arXiv240614466S}
}

@ARTICLE{Miller2015,
       author = {{Miller}, M. Coleman and {Lamb}, Frederick K.},
        title = "{Determining Neutron Star Properties by Fitting Oblate-star Waveform Models to X-Ray Burst Oscillations}",
      journal = {\apj},
         year = 2015,
        month = jul,
       volume = {808},
       number = {1},
          eid = {31},
        pages = {31},
          doi = {10.1088/0004-637X/808/1/31},
archivePrefix = {arXiv},
       eprint = {1407.2579},
 primaryClass = {astro-ph.HE},
       adsurl = {https://ui.adsabs.harvard.edu/abs/2015ApJ...808...31M}
}

@ARTICLE{Salmi2023,
       author = {{Salmi}, Tuomo and {Vinciguerra}, Serena and {Choudhury}, Devarshi and {Watts}, Anna L. and {Ho}, Wynn C.~G. and {Guillot}, Sebastien and {Kini}, Yves and {Dorsman}, Bas and {Morsink}, Sharon M. and {Bogdanov}, Slavko},
        title = "{Atmospheric Effects on Neutron Star Parameter Constraints with NICER}",
      journal = {\apj},
         year = 2023,
        month = oct,
       volume = {956},
       number = {2},
          eid = {138},
        pages = {138},
          doi = {10.3847/1538-4357/acf49d},
archivePrefix = {arXiv},
       eprint = {2308.09319},
 primaryClass = {astro-ph.HE},
       adsurl = {https://ui.adsabs.harvard.edu/abs/2023ApJ...956..138S}
}

@ARTICLE{Pout1,
       author = {{Poutanen}, Juri and {Beloborodov}, Andrei M.},
        title = "{Pulse profiles of millisecond pulsars and their Fourier amplitudes}",
      journal = {\mnras},
         year = 2006,
        month = dec,
       volume = {373},
       number = {2},
        pages = {836-844},
          doi = {10.1111/j.1365-2966.2006.11088.x},
archivePrefix = {arXiv},
       eprint = {astro-ph/0608663},
 primaryClass = {astro-ph},
       adsurl = {https://ui.adsabs.harvard.edu/abs/2006MNRAS.373..836P}
}

@ARTICLE{Beloborodov2002,
       author = {{Beloborodov}, Andrei M.},
        title = "{Gravitational Bending of Light Near Compact Objects}",
      journal = {\apjl},
         year = 2002,
        month = feb,
       volume = {566},
       number = {2},
        pages = {L85-L88},
          doi = {10.1086/339511},
archivePrefix = {arXiv},
       eprint = {astro-ph/0201117},
 primaryClass = {astro-ph},
       adsurl = {https://ui.adsabs.harvard.edu/abs/2002ApJ...566L..85B}
}

@ARTICLE{Baubock2019,
       author = {{Baub{\"o}ck}, Michi and {Psaltis}, Dimitrios and {{\"O}zel}, Feryal},
        title = "{Atmospheric Structure and Radiation Pattern for Neutron-star Polar Caps Heated by Magnetospheric Return Currents}",
      journal = {\apj},
         year = 2019,
        month = feb,
       volume = {872},
       number = {2},
          eid = {162},
        pages = {162},
          doi = {10.3847/1538-4357/aafe08},
archivePrefix = {arXiv},
       eprint = {1901.01274},
 primaryClass = {astro-ph.HE},
       adsurl = {https://ui.adsabs.harvard.edu/abs/2019ApJ...872..162B}
}

@ARTICLE{Miller2019,
       author = {{Miller}, M.~C. and {Lamb}, F.~K. and {Dittmann}, A.~J. and {Bogdanov}, S. and {Arzoumanian}, Z. and {Gendreau}, K.~C. and {Guillot}, S. and {Harding}, A.~K. and {Ho}, W.~C.~G. and {Lattimer}, J.~M. and {Ludlam}, R.~M. and {Mahmoodifar}, S. and {Morsink}, S.~M. and {Ray}, P.~S. and {Strohmayer}, T.~E. and {Wood}, K.~S. and {Enoto}, T. and {Foster}, R. and {Okajima}, T. and {Prigozhin}, G. and {Soong}, Y.},
        title = "{PSR J0030+0451 Mass and Radius from NICER Data and Implications for the Properties of Neutron Star Matter}",
      journal = {\apjl},
         year = 2019,
        month = dec,
       volume = {887},
       number = {1},
          eid = {L24},
        pages = {L24},
          doi = {10.3847/2041-8213/ab50c5},
archivePrefix = {arXiv},
       eprint = {1912.05705},
 primaryClass = {astro-ph.HE},
       adsurl = {https://ui.adsabs.harvard.edu/abs/2019ApJ...887L..24M}
}

@ARTICLE{Miller2021,
       author = {{Miller}, M.~C. and {Lamb}, F.~K. and {Dittmann}, A.~J. and {Bogdanov}, S. and {Arzoumanian}, Z. and {Gendreau}, K.~C. and {Guillot}, S. and {Ho}, W.~C.~G. and {Lattimer}, J.~M. and {Loewenstein}, M. and {Morsink}, S.~M. and {Ray}, P.~S. and {Wolff}, M.~T. and {Baker}, C.~L. and {Cazeau}, T. and {Manthripragada}, S. and {Markwardt}, C.~B. and {Okajima}, T. and {Pollard}, S. and {Cognard}, I. and {Cromartie}, H.~T. and {Fonseca}, E. and {Guillemot}, L. and {Kerr}, M. and {Parthasarathy}, A. and {Pennucci}, T.~T. and {Ransom}, S. and {Stairs}, I.},
        title = "{The Radius of PSR J0740+6620 from NICER and XMM-Newton Data}",
      journal = {\apjl},
         year = 2021,
        month = sep,
       volume = {918},
       number = {2},
          eid = {L28},
        pages = {L28},
          doi = {10.3847/2041-8213/ac089b},
archivePrefix = {arXiv},
       eprint = {2105.06979},
 primaryClass = {astro-ph.HE},
       adsurl = {https://ui.adsabs.harvard.edu/abs/2021ApJ...918L..28M}
}

@ARTICLE{Riley2019,
       author = {{Riley}, T.~E. and {Watts}, A.~L. and {Bogdanov}, S. and {Ray}, P.~S. and {Ludlam}, R.~M. and {Guillot}, S. and {Arzoumanian}, Z. and {Baker}, C.~L. and {Bilous}, A.~V. and {Chakrabarty}, D. and {Gendreau}, K.~C. and {Harding}, A.~K. and {Ho}, W.~C.~G. and {Lattimer}, J.~M. and {Morsink}, S.~M. and {Strohmayer}, T.~E.},
        title = "{A NICER View of PSR J0030+0451: Millisecond Pulsar Parameter Estimation}",
      journal = {\apjl},
         year = 2019,
        month = dec,
       volume = {887},
       number = {1},
          eid = {L21},
        pages = {L21},
          doi = {10.3847/2041-8213/ab481c},
archivePrefix = {arXiv},
       eprint = {1912.05702},
 primaryClass = {astro-ph.HE},
       adsurl = {https://ui.adsabs.harvard.edu/abs/2019ApJ...887L..21R}
}

@ARTICLE{Riley2021,
       author = {{Riley}, Thomas E. and {Watts}, Anna L. and {Ray}, Paul S. and {Bogdanov}, Slavko and {Guillot}, Sebastien and {Morsink}, Sharon M. and {Bilous}, Anna V. and {Arzoumanian}, Zaven and {Choudhury}, Devarshi and {Deneva}, Julia S. and {Gendreau}, Keith C. and {Harding}, Alice K. and {Ho}, Wynn C.~G. and {Lattimer}, James M. and {Loewenstein}, Michael and {Ludlam}, Renee M. and {Markwardt}, Craig B. and {Okajima}, Takashi and {Prescod-Weinstein}, Chanda and {Remillard}, Ronald A. and {Wolff}, Michael T. and {Fonseca}, Emmanuel and {Cromartie}, H. Thankful and {Kerr}, Matthew and {Pennucci}, Timothy T. and {Parthasarathy}, Aditya and {Ransom}, Scott and {Stairs}, Ingrid and {Guillemot}, Lucas and {Cognard}, Ismael},
        title = "{A NICER View of the Massive Pulsar PSR J0740+6620 Informed by Radio Timing and XMM-Newton Spectroscopy}",
      journal = {\apjl},
         year = 2021,
        month = sep,
       volume = {918},
       number = {2},
          eid = {L27},
        pages = {L27},
          doi = {10.3847/2041-8213/ac0a81},
archivePrefix = {arXiv},
       eprint = {2105.06980},
 primaryClass = {astro-ph.HE},
       adsurl = {https://ui.adsabs.harvard.edu/abs/2021ApJ...918L..27R}
}

@ARTICLE{Suleimanov2012,
       author = {{Suleimanov}, V. and {Poutanen}, J. and {Werner}, K.},
        title = "{X-ray bursting neutron star atmosphere models using an exact relativistic kinetic equation for Compton scattering}",
      journal = {\aap},
         year = 2012,
        month = sep,
       volume = {545},
          eid = {A120},
        pages = {A120},
          doi = {10.1051/0004-6361/201219480},
archivePrefix = {arXiv},
       eprint = {1208.1467},
 primaryClass = {astro-ph.HE},
       adsurl = {https://ui.adsabs.harvard.edu/abs/2012A&A...545A.120S}
}

@ARTICLE{Suleimanov2020,
       author = {{Suleimanov}, Valery F. and {Poutanen}, Juri and {Werner}, Klaus},
        title = "{Observational appearance of rapidly rotating neutron stars. X-ray bursts, cooling tail method, and radius determination}",
      journal = {\aap},
         year = 2020,
        month = jul,
       volume = {639},
          eid = {A33},
        pages = {A33},
          doi = {10.1051/0004-6361/202037502},
archivePrefix = {arXiv},
       eprint = {2005.09759},
 primaryClass = {astro-ph.HE},
       adsurl = {https://ui.adsabs.harvard.edu/abs/2020A&A...639A..33S}
}

@ARTICLE{Zhou2026,
       author = {{Zhou}, Tianzhe and {Huang}, Chun},
        title = "{GPU-accelerated X-ray pulse profile modeling}",
      journal = {\aap},
         year = 2026,
        month = may,
       volume = {709},
          eid = {A111},
        pages = {A111},
          doi = {10.1051/0004-6361/202659305},
archivePrefix = {arXiv},
       eprint = {2510.07764},
 primaryClass = {astro-ph.HE},
       adsurl = {https://ui.adsabs.harvard.edu/abs/2026A&A...709A.111Z}
}

@ARTICLE{atmosphere,
       author = {{Salmi}, Tuomo and {Suleimanov}, Valery F. and {N{\"a}ttil{\"a}}, Joonas and {Poutanen}, Juri},
        title = "{Magnetospheric return-current-heated atmospheres of rotation-powered millisecond pulsars}",
      journal = {\aap},
         year = 2020,
        month = sep,
       volume = {641},
          eid = {A15},
        pages = {A15},
          doi = {10.1051/0004-6361/202037824},
archivePrefix = {arXiv},
       eprint = {2002.11427},
 primaryClass = {astro-ph.HE},
       adsurl = {https://ui.adsabs.harvard.edu/abs/2020A&A...641A..15S}
}
\end{document}